\documentclass{jnmp}
\usepackage{amsmath}

\def\Beq{\begin{gather}}
\def\Eeq{\end{gather}}
\def\Beqa{\begin{eqnarray}}
\def\Eeqa{\end{eqnarray}}

\def\CC{{\bf C}}
\def\Al{{\lambda_2}}
\def\Bl{{\lambda_1}}
\def\Cl{{\lambda_0}}

\def\CC{{\bf C}}

\def\ee{{\rm{e}}}

\JNMPnumberwithin{equation}{section}

\theoremstyle{definition}

\begin{document}

%
\renewcommand{\evenhead}{S. Matsutani}
\renewcommand{\oddhead}{Toda Equations and $\sigma$-Functions
of Genera One and Two}

%
\thispagestyle{empty}

\FirstPageHead{*}{*}{20**}{\pageref{firstpage}--\pageref{lastpage}}{Article}

\copyrightnote{200*}{S. Matsutani}

\Name{ Toda Equations and $\sigma$-Functions of Genera One and Two}

\label{firstpage}

\Author{Shigeki Matsutani}

\Address{8-21-1, Higashi-Linkan, Sagamihara 228-0811, JAPAN\\
~~E-mail: RXB01142@nifty.com\\[10pt]}

\Date{Received Month *, 200*; Revised Month *, 200*;
Accepted Month *, 200*}

\begin{abstract}
\noindent
We study the Toda equations in the
continuous level, discrete level
and ultradiscrete level
in terms of elliptic and hyperelliptic $\sigma$
and $\psi$ functions of genera one and two.
The ultradiscrete Toda equation appears as
a discrete-valuation of recursion relations of $\psi$
functions.
\end{abstract}

%
\section{Introduction}

Recently Kimijima and Tokihiro found
solutions of the discrete and
ultradiscrete Toda equations
in terms of elliptic and hyperelliptic $\theta$ functions
\cite{tokihiro}. In this article
we present another type of solution of another type of
the discretization of the Toda equation \cite{hirota}
and its ultradiscretization \cite{takahashi}
associated with algebraic curves of genera one and two.

Elliptic and hyperelliptic
$\sigma$ functions are related to nonlinear differential
equations from the beginning \cite{Baker2, Klein, Weierstrass}.
We study the Toda equations
at the continuous level, discrete level and ultradiscrete
level in terms of $\sigma$ functions
and $\psi$ functions of genera one and two. We show that
these equations have solutions expressed in terms of $\sigma$
and $\psi$ functions. Here
the $\psi$ functions are defined by rational
functions of the $\sigma$ functions.

In \cite{matsu2} it was shown that the $\psi$ functions
can be related to discrete nonlinear equations, such
as the discrete Painlev\'e equations.
This article can be considered as one of a series in which
relations between $\psi$ functions and discrete nonlinear
difference equations are unfolded.

Further, as was mentioned in \cite{matsu3}, the ultradiscrete sometimes
can be regarded as a valuation of a related field.
This article shows that, in the case of the Toda equation,
it can be also realized as
a discrete valuation of a function field over a Jacobi variety.

In Section 2 we concentrate on the genus one case
and give concrete solutions.
We investigate the genus two version in Section 3.
It is shown that
all solutions of the Toda equations
in this study are connected with the
addition formulae of the $\sigma$ functions.

\section{Genus One Case}

In this Section we deal with an elliptic
curve given by
\begin{eqnarray}
C_1\ :\
\frac{1}{4} \bar y^2 = y^2 &=&x^3 + \lambda_2 x^2
+\lambda_1 x + \lambda_0\nonumber \\
&=& (x-b_1)(x-b_2)(x-b_3),
\label{eq:ecurve}
\end{eqnarray}
where the $b$ s are complex numbers.

\subsection{Continuous Toda Equation}

Firstly we give a $\wp$ function solution of
the continuous Toda equation \cite{toda}.
We treat the Weierstrass elliptic $\sigma$ function
associated with the curve $C_1$, which is
connected with the Weierstrass $\wp$ function
by
\begin{equation}
\wp(u) =- \frac{d^2}{d u^2} \log \sigma(u).
\end{equation}
A local parameter $u$ in $C_1$ is given by
\begin{equation}
u = \int^{(x,y)}_\infty \frac{d x}{2 y},
\end{equation}
and $\wp(u)$ is equal to $x(u)$. Here $\infty$
is the infinity point of $C_1$.

The addition formula of the $\sigma$ functions is
given by
\begin{equation}
-[\wp(u) - \wp(v)] =
\frac{\sigma(v+u)\sigma(u-v)}{[\sigma(v)\sigma(u)]^2}.
\label{eq:add1}
\end{equation}
By differentiating the logarithm of (\ref{eq:add1}) by $u$ twice,
we have
\begin{equation}
- \frac{d^2}{d u^2} \log [\wp(u)-\wp(v)]
= \wp(u+v) - 2\wp(u) + \wp(u-v).
\end{equation}
For a constant number $u_0$,
by letting $u=n u_0+t$, $v=u_0$ and $b:=\wp(u_0)$, we have
\begin{eqnarray}
&&-\frac{d^2}{d t^2} \log [\wp(n u_0+t)-b]
= [\wp((n+1)u_0+t)-b] \nonumber\\
&&\quad- 2[\wp(n u_0+t)-b]
+ [\wp((n-1)u_0+t)-b].
\end{eqnarray}
Further by letting
$q_n := \log [\wp((n+1)u_0+t)-b]$,
we have
\begin{equation}
-\frac{d^2}{d t^2}(q_n) = \ee^{q_{n+1}}-2 \ee^{q_n} +\ee^{q_{n-1}}.
\end{equation}
This is identified with the continuous Toda lattice equation.
In fact, by letting $q_n = Q_n - Q_{n-1}$,
$Q_n$ obeys the nonlinear differential equation of a nonlinear lattice
\cite{toda}.
It is clear that this elliptic solution comes from the
addition formula (\ref{eq:add1}).
Later we show that a genus two solution of
the Toda equation
can be expressed by a similar form in \S 3-1.

\subsection{{ Discrete Toda Equation and $\psi$ Functions}}

Though there are several models of the discrete Toda equations,
we concentrate on a model given in \cite{hirota}.
In this subsection we give elliptic solutions of
the discrete Toda equation.

The elliptic $\psi$ function is given by
\begin{equation}
\psi_n(u)=
\frac{\sigma(nu)}{\sigma(u)^{n^2}} \label{eq:psi1}
\end{equation}
and, due to the
addition formula (\ref{eq:add1}),
it satisfies the recursion relation \cite{weber},
\begin{equation}
\psi_{n+m} \psi_{m-n}
=\left| \begin{array}{r r}
\psi_{m-1}\psi_n & \psi_{m}\psi_{n+1} \\
\psi_{m}\psi_{n-1} & \psi_{m+1}\psi_{n}
\end{array} \right|. \label{eq:rec_psi1}
\end{equation}
Further $\psi_n$ can also be computed using
the Brioches-Kiepert relation
\cite{Biroschi, Kie},
\begin{equation}
\psi_n(u)=
\frac{ (-1)^{n-1}}{[1!2!\cdots (n-1)!]^2}
\left|\begin{array}{r r r r}
\wp'(u) & \wp''(u) & \cdots & \wp^{(n-1)}(u) \\
\wp''(u) & \wp'''(u) & \cdots & \wp^{(n)}(u) \\
\vdots & \vdots & \ddots & \vdots \\
\wp^{(n-1)}(u) & \wp^{(n)}(u) & \cdots & \wp^{(2n-3)}(u) \\
\end{array}\right|.\label{eq:BK1}
\end{equation}
Here derivatives of $u$ are denoted by $'$ and $^{(n)}$.
By noting
$ \frac{d}{d u} = 2y \frac{d }{d x}$,
the $\psi$ function is a polynomial of $x$ and $y$ over the
complex number $\bf C$.
In fact the $\psi$ function can be explicitly obtained as,
\begin{eqnarray}
\psi_1(u)&=&1, \nonumber \\
\psi_2(u)&=&-2y, \nonumber \\
\psi_3(u)&=&3 x^4 +4\Al x^3 +6 \Bl x^2 +12\Cl x-\Bl^2+4\Al \Cl,
\nonumber \\
\psi_4(u)&=&-4y[ x^6+2\Al x^5+5 \Bl x^4 +20\Cl x^3\nonumber\\
& &+(20\Al\Cl-5\Bl^2)x^2+(8\Al^2\Cl-2\Al\Bl^2-4\Bl\Cl)x
\nonumber\\
& & +4\Al\Bl\Cl -\Bl^3-8\Cl^2].
\end{eqnarray}
For $\psi_n$, $n>4$, we have the recursion relations
\begin{eqnarray}
\psi_{2n+1}&=&\psi_{n+2}\psi_n^3-\psi_{n+1}^3\psi_{n-1},
\nonumber \\
\psi_{2n}&=&\psi_n(
\psi_{n+2}\psi_{n-1}^2-\psi_{n+1}^2\psi_{n-2})/\psi_2.
\label{eq:rec4}
\end{eqnarray}

Thus we know that
\begin{eqnarray}
\psi_n(u)&\in&{\bf C}[x,\lambda_0,\lambda_1,\lambda_2]
\quad \mbox{for odd } n,\nonumber \\
\psi_n(u)&\in&{\bf C}[x,\lambda_0,\lambda_1,\lambda_2]y
\quad \mbox{for even } n .\label{eq:psipoly}
\end{eqnarray}
When $n$ is odd, $\psi_n$ is a polynomial of $x$ whose
order is $(n^2-1)/2$
and, for an even $n$, the order of $x$ for $\psi_n/y$ is
$(n^2-4)/2$.
For specific curves, we give explicit forms of $\psi_n$ in the
Appendix.

We comment on the properties of $\psi$ functions.
We note that $\sigma(u)$ is characterized by the property
that it has no singularity with respect to $u \in {\bf C}$
and its zeros are identified with a
lattice points generated by the periodicity $2\omega$ of $\wp(u)$,
$\wp(u+2\omega)=\wp(u)$.
In other words the zero of $\sigma$
is congruent to the origin of the local parameter $u$ modulo the lattice.
Accordingly,
as $\psi$ function is a function over the curve $C$,
we conclude that
a point satisfying
\begin{equation}
\psi_n(u)=0
\end{equation}
is a point for which $nu$ is equal to the lattice point again.
In other words we have $n$-cyclic points as
zeros of $\psi_n$.
Conversely it can be shown that
the polynomial of $x$ and $y$ whose zeros
multiplied by $n$ are lattice points
must be $\psi_n$ modulo constant factors.

Hence, if $n$ is a factor of $m$, {\it i.e.},
$n|m$, it is clear that $\psi_n$ is divided by $\psi_m$,
\begin{equation}
\psi_n | \psi_m.
\end{equation}

For mutually coprime numbers $p, q$ and an integer $n_0$,
we introduce
\begin{equation}
\phi_i^{\ j} := \psi_{n_0 + pi +q j}.
\end{equation}
By letting $n\equiv n_0+pi +q j$ we have
\begin{eqnarray}
\psi_{n+p}\psi_{n-p}
&=& \psi_n^2 \psi_{p+1}\psi_{p-1} - \psi_{p}^2 \psi_{n+1}\psi_{n-1},
\nonumber \\
\psi_{n+q}\psi_{n-q}
&=&\psi_n^2 \psi_{q+1}\psi_{q-1} - \psi_{q}^2 \psi_{n+1}\psi_{n-1}.
\end{eqnarray}
The components $\psi_{n+1}\psi_{n-1}$ in both formulae
give a relation, viz
\begin{equation}
(\psi_{n+p}\psi_{n-p}
- \psi_n^2 \psi_{p+1}\psi_{p-1})\psi_{q}^2
=
(\psi_{n+q}\psi_{n-q}
- \psi_n^2 \psi_{q+1}\psi_{q-1}) \psi_{p}^2.\label{eq:rell}
\end{equation}
Noting $n\equiv n_0+pi +q j$ (\ref{eq:rell}) can be regarded
as an evolution equation for $i$ and $j$ when
we consider $\psi_{q}$, $\psi_{p}$ and $\psi_{q\pm1}$ as
initial values.
Let us assume that $\psi_q$ $\psi_p$ and $\psi_{p\pm q}$
are not equal to zero by choosing
the parameter $u$.
By letting $\delta:=\psi_q/\psi_p$ and
$c(1-\delta):=-\psi_{p+q}\psi_{p-q}/(\psi_q)^2$
we have
\begin{equation}
\phi_i^{\ j+1}\phi_i^{\ j-1}
-c(1-\delta^2) \phi_i^{\ j}\phi_i^{\ j}
-\delta^2\phi_{i+1}^{\ \ j}\phi_{i-1}^{\ \ j}=0.
\label{eq:dtoda1}
\end{equation}

For later convenience we do not fix $c$ and
define
\begin{equation}
V_i^{\ j}:=\left(\frac{\phi_{i+1}^{\ \ j}\phi_{i-1}^{\ \ j}}
{\phi_{i}^{\ j}\phi_{i}^{\ j}} \right)-c.
\label{eq:vij}
\end{equation}
Then we obtain
\begin{equation}
\log\left(\frac{(c+V_i^{\ j})^2}
{(c+V_{i}^{\ j+1})(c+V_{i}^{\ j-1})}\right)
=\log\left(\frac{(c+\delta^2 V_i^{\ j})^2}
{(c+\delta^2 V_{i+1}^{\ j})(c+\delta^2 V_{i-1}^{\ j})}
\right).
\label{eq:dtodaV}
\end{equation}
When $c=1$,
this equation is one of discrete versions of the Toda equation,
which appeared in \cite{hirota}.

The condition $c=1$ means that
\begin{equation}
\psi_{p+q}\psi_{p-q} + \psi_p\psi_p-\psi_q\psi_q=0,
\label{eq:cond}
\end{equation}
which is an equation of $(p^2+q^2-2)$-order with respect to $x$.
In other words, for a point $u$ satisfying (\ref{eq:cond}),
we have a solution of the discrete Toda equation
in \cite{hirota}
in terms of $\psi$ functions.

Further we introduce
$U_i^{\ j}:=(V_i^{\ j}+c)$ which satisfies
\begin{equation}
\left(\frac{(U_i^{\ j})^2}
{(U_{i}^{\ j+1})(U_{i}^{\ j-1})}\right)
=\left(\frac{(c(1-\delta^2)+\delta^2 U_i^{\ j})^2}
{(c(1-\delta^2)+\delta^2 U_{i+1}^{\ j})
(c(1-\delta^2)+\delta^2 U_{i-1}^{\ j})}
\right). \label{eq:dtoda2}
\end{equation}
We investigate this equation with general $c$ and
go on to call it the discrete Toda equation in this article.
We note that this solution is due to the
recursion relation (\ref{eq:rec_psi1})
which comes from the addition formula (\ref{eq:add1}).

\subsection{{ Periodic Solutions of Discrete Toda Equation}}

For general $c$ in (\ref{eq:vij}) we consider a periodic
solution of (\ref{eq:dtoda2}).
It is obvious that, when $\psi_n=0$,
$\psi_{nr}=0$. Thus there may exist a point, $u_1$,
such that
\begin{equation}
\psi_i(u_1) = \psi_{n+i}(u_1).
\end{equation}
In fact we have solutions of (\ref{eq:dtoda2}) for a curve
$y^2 = x^2(x+1/4)$ and its point
$x=-1$ (See the Appendix). The $\psi$ function
has values as in Table 1.
$$
\centerline{
\vbox{
\baselineskip =10pt
\tabskip = 1em
\halign{&\hfil#\hfil \cr
\multispan7 \hfil Table 1: $\psi_n$ at $x = -1$ \hfil \cr
\noalign{\smallskip}
\noalign{\hrule height0.8pt}
\noalign{\smallskip}
& $n$ &\strut\vrule& 0 & 1 & 2 & 3 & 4 & 5 & 6 & 7 & 8 & 9 & 10 & 11 &
12 \cr
\noalign{\smallskip}
\noalign{\hrule height0.3pt}
\noalign{\smallskip}
& $\psi_n$ & \strut\vrule& 0 & 1 & $-\sqrt{-3}$ & 2 & $-\sqrt{-3}$ &
1 & 0 & $-1$ & $\sqrt{-3}$ & $-2$ & $\sqrt{-3}$ & $-1$ & 0 \cr
\noalign{\smallskip}
\noalign{\hrule height0.8pt}
}
}
}
$$

For $(p,q)=(3,2)$ and $n_0=0$, {\it i.e.},
$\delta^2 = -3/4$ and $c(1-\delta^2)=1/4$,
we have a periodic solution of (\ref{eq:dtoda2}).
$$
\centerline{
\vbox{
\baselineskip =10pt
\tabskip = 1em
\halign{&\hfil#\hfil \cr
\multispan7 \hfil Table 2: $U_i^j$ ($p=3$, $q=2$ case)
\hfil \cr
\noalign{\smallskip}
\noalign{\hrule height0.8pt}
\noalign{\smallskip}
& $j\backslash i$ &\strut\vrule& 0 & 1 & 2 & 3 \cr
\noalign{\smallskip}
\noalign{\hrule height0.3pt}
\noalign{\smallskip}
& $0$ & \strut\vrule& $\infty$ & 0 & $\infty$ & 0 \cr
& $1$ & \strut\vrule& $1/3$ & 3 & 1/3 & 3 \cr
& $2$ & \strut\vrule& $1/3$ & 3 & 1/3 & 3 \cr
& $3$ & \strut\vrule& $\infty$ & 0 & $\infty$ & 0 \cr
\noalign{\smallskip}
\noalign{\hrule height0.8pt}
}
}
}
$$

For $(p,q)=(2,3)$ and $n_0=0$, {\it i.e.},
$\delta^2 = -4/3$ and $c(1-\delta^2)=1/3$,
another periodic solution of (\ref{eq:dtoda2})
is given in Table 3.
$$
\centerline{
\vbox{
\baselineskip =10pt
\tabskip = 1em
\halign{&\hfil#\hfil \cr
\multispan7 \hfil Table 3: $U_i^j$ ($p=2$, $q=3$ case) \hfil \cr
\noalign{\smallskip}
\noalign{\hrule height0.8pt}
\noalign{\smallskip}
& $j\backslash i$ &\strut\vrule& 0 & 1 & 2 & 3 \cr
\noalign{\smallskip}
\noalign{\hrule height0.3pt}
\noalign{\smallskip}
& $0$ & \strut\vrule& $\infty$ & 0 & 0 & $\infty$ \cr
& $1$ & \strut\vrule& $1/4$ & $-2$ & $-2$ & 1/4 \cr
& $2$ & \strut\vrule& $\infty$ & 0 & 0 & $\infty$ \cr
& $3$ & \strut\vrule& $1/4$ & $-2$ & $-2$ & 1/4 \cr
\noalign{\smallskip}
\noalign{\hrule height0.8pt}
}
}
}
$$

\subsection{{
Ultradiscrete Toda Equations}}

In this subsection we investigate the ultradiscrete version
of the Toda equation using $\psi$ functions.

For the elliptic curve $C_1$ a local parameter $t$
should be characterized by
\begin{equation}
\begin{array}{lcc}
\mbox{for a generic point $x_0$ in $C_1$,} &:& t = x-x_0,\\
\mbox{for a finite branch point $b_i$ in $C_1$, } &:& t = \sqrt{x-b_i},\\
\mbox{for the infinity point $\infty$ in $C_1$, } &:& t = 1/\sqrt{x}.
\end{array}
\end{equation}
Let a localization of the commutative ring $R=\CC[x,y]/(y^2-f(x))$
at $u_0$ be denoted by $R_{u_0}$.
Let $K_{u_0}$ be a field of Laurent transformations at $u_0$ of rational
functions.
The valuation of the field $K_{u_0}$ is given that
for $f\in K_{u_0}$, let $\mbox{val}(f) =\infty$ if $f=0$,
if $f$ is given by
\begin{equation}
f(u)=a(u - u_0)^m + {\cal O}((u - u_0)^{m+1})
\end{equation}
for $a\neq 0$, let $\mbox{val}(f) = m$ \cite{Ha}.
Denoting set of integers by $\bf Z$,
the discrete valuation is known as a map
\begin{equation}
\mbox{val}:K_{u_0} \to \bf Z + \infty,
\end{equation}
which satisfies
\begin{eqnarray}
\mbox{val}(f g) &=& \mbox{val}(f) + \mbox{val}(g),\nonumber \\
\mbox{val}(f+g) &\ge& \min(\mbox{val}(f) , \mbox{val}(g))
\label{eq:val2}.
\end{eqnarray}
For example the inequality in (\ref{eq:val2})
appears due to a case,
$k=m$ and $a=-b$
for $f=a(u - u_0)^m + \cdots$ and $g=b(u - u_0)^k + \cdots$
with $(a,b\neq0)$. Inversely, as long as we avoid
such a case, we can regard the second relation in (\ref{eq:val2})
as an equality.

$R_{u_0}$ can be expressed as
\begin{equation}
R_{u_0} = \{ \ f \in K_{u_0} \ | \ \mbox{val}(f) \ge 0 \ \}.
\end{equation}
$
R_{u_0}^{\times} := \{ \ f \in K_{u_0} \ | \ \mbox{val}(f) = 0 \ \}
$
is a multiplication group in $R_{u_0}$. An element in
$R_{u_0}^{\times}$ is called unit. The subset
$
{\bf {m}} := \{ \ f \in K_{u_0} \ | \ \mbox{val}(f) >0 \ \}
$
of $R_{u_0}$
is a unique maximal ideal in $R_{u_0} $ and thus we have
a filter structure,
\begin{equation}
{\bf {m}}^k \supset {\bf {m}}^{k+1}.
\end{equation}
Here the multiplication among ideals is given as a set of sum of
multiplications of elements in the ideals.
Due to the filter structure there naturally appears
a nonarchimedean distance given by
\begin{equation}
|f - g|_{\mbox{\small val}} := \exp( - \mbox{val}( f - g ) ).
\end{equation}
Thus an element $f$ in ${\bf {m}}$ is a smaller element than unity,
{\it i.e.}, $|f|_{\mbox{\small val}}<1$.
When $\delta$ behaves like a small parameter \cite{hirota},
we regard it as an
element of $\bf m$, {\it i.e.},
\begin{equation}
\delta \equiv \frac{\psi_q}{\psi_p}(u)\in { \bf m}.
\end{equation}

We now consider the point satisfying
\begin{equation}
c(1-\delta^2) =\frac{\psi_{p+q}(u)\psi_{p-q}(u)}
{\psi_{p}(u)^2}\in R_{u}^\times.
\end{equation}
Define
\begin{equation}
f_i^{\ j}:=-\mbox{val}(U_i^{ j}), \quad
d :=- \mbox{val}(\delta^2).
\end{equation}
When we expand them as
$ U_i^{ j} \delta^{2}=a(u - u_0)^m + \cdots$ and
$c(1-\delta^2)=b(u - u_0)^k + \cdots$ with
$a,b\neq0$, we assume that
for any $i$ and $j$, $k$ is not equal to $m$
or $a$ is not equal to $-b$ if $k=m$.
Then (\ref{eq:dtoda2}) becomes
\begin{equation}
f_{i}^{\ j+1} - 2f_{i}^{\ j}+f_{i}^{\ j-1}
=\mbox{max}(0, f_{i+1}^{\ j}+d)-2
\mbox{max}(0, f_{i}^{\ j}+d)
+\mbox{max}(0,f_{i-1}^{\ j}+d ).
\label{eq:uDTE1}
\end{equation}
This is identified with the ultradiscrete Toda equation in \cite{takahashi}.

Let us consider solutions of
the ultradiscrete Toda equation (\ref{eq:uDTE1}).
By letting
$g_n:=\mbox{val}(\psi_n)$,
\begin{equation}
f_{i}^{\ j} = g_{i+1}^{\ j}-2g_{i}^{\ j}+g_{i-1}^{\ j}.
\end{equation}
For the curve $y^2 = x^3 +1/4$, and
$u_0$ at $x(u_0)=(-1/4)^{1/3}$, we have $g_{i}$ as in Table 4:
$$
\centerline{
\vbox{
\baselineskip =10pt
\tabskip = 1em
\halign{&\hfil#\hfil \cr
\multispan7 \hfil Table 4: $g_n$ at $y=0$ \hfil \cr
\noalign{\smallskip}
\noalign{\hrule height0.8pt}
\noalign{\smallskip}
& $n$ &\strut\vrule& 0 & 1 & 2 & 3 & 4 & 5 & 6 & 7 & 8 & 9 & 10 & 11 &
12 &$\cdots$\cr
\noalign{\smallskip}
\noalign{\hrule height0.3pt}
\noalign{\smallskip}
& $g_n$ & \strut\vrule& $\infty$ & 0 & 1 & 0 & 1 & 0 &
1 & 0 & 1 & 0 & 1 & 0 & 1 &$\cdots$\cr
\noalign{\smallskip}
\noalign{\hrule height0.8pt}
}
}
}
$$
Then we have a solution of (\ref{eq:uDTE1}) for
$p=3$, $q=2$ and $n_0=0$: $d=-2$, $\mbox{val}(c(1-\delta^2)=0$,
$$
\centerline{
\vbox{
\baselineskip =10pt
\tabskip = 1em
\halign{&\hfil#\hfil \cr
\multispan7 \hfil Table 5: $f_i^j$ ($p=3$, $q=2$ case)
\hfil \cr
\noalign{\smallskip}
\noalign{\hrule height0.8pt}
\noalign{\smallskip}
& $j\backslash i$ &\strut\vrule& 1 & 2 & 3 & 4 & 5 & $\cdots$\cr
\noalign{\smallskip}
\noalign{\hrule height0.3pt}
\noalign{\smallskip}
& $0$ & \strut\vrule& $\infty$ & $-2$ & 2 & $-2$ & 2 & $\cdots$ \cr
& $1$ & \strut\vrule& 2 & $-2$ & 2 & $-2$ & 2 & $\cdots$\cr
& $2$ & \strut\vrule& 2 & $-2$ & 2 & $-2$ & 2 & $\cdots$\cr
& $3$ & \strut\vrule& 2 & $-2$ & 2 & $-2$ & 2 & $\cdots$\cr
& $\vdots$ & \strut\vrule& $\vdots$& $\vdots$& $\vdots$ & $\vdots$&
$\vdots$& $\ddots$\cr
\noalign{\smallskip}
\noalign{\hrule height0.8pt}
}
}
}
$$

Next we deal with a curve $y^2 = x^3 -x$ and a point
$u_0(x=0)$. The values of $g_i$ are given in Table 6.
$$
\centerline{
\vbox{
\baselineskip =10pt
\tabskip = 1em
\halign{&\hfil#\hfil \cr
\multispan7 \hfil Table 6: $g_n$ at $x=0$ \hfil \cr
\noalign{\smallskip}
\noalign{\hrule height0.8pt}
\noalign{\smallskip}
& $n$ &\strut\vrule& 0 & 1 & 2 & 3 & 4 & 5 & 6 & 7 & 8 & 9 & 10 & 11 & 12 &
$\cdots$\cr
\noalign{\smallskip}
\noalign{\hrule height0.3pt}
\noalign{\smallskip}
& $g_n$ & \strut\vrule& $\infty$ & 0 & 1 & 4 & 5 & 8 &
13& 16 & 21 & 28 & 33 & 40 & 49 &$\cdots$\cr
\noalign{\smallskip}
\noalign{\hrule height0.8pt}
}
}
}
$$
When $(p,q, n_0)=(5,2,0)$, we have $\mbox{val}(c(1-\delta^2))=0$,
$d=14$ and Table 7.
$$
\centerline{
\vbox{
\baselineskip =10pt
\tabskip = 1em
\halign{&\hfil#\hfil \cr
\multispan7 \hfil Table 7: $f_i^j$ ($p=5$, $q=2$ case)
\hfil \cr
\noalign{\smallskip}
\noalign{\hrule height0.8pt}
\noalign{\smallskip}
& $j\backslash i$ &\strut\vrule& 1 & 2 & 3 & 4 & $\cdots$\cr
\noalign{\smallskip}
\noalign{\hrule height0.3pt}
\noalign{\smallskip}
& $0$ & \strut\vrule& $\infty$ & $18$ &14 & $18$ & $\cdots$ \cr
& $1$ & \strut\vrule& 18 & $14$ &18 & $18$ & $\cdots$\cr
& $2$ & \strut\vrule& 14 & $18$ &18 & $14$ &$\cdots$\cr
& $\vdots$ & \strut\vrule& $\vdots$& $\vdots$& $\vdots$ &
$\vdots$ &$\ddots$\cr
\noalign{\smallskip}
\noalign{\hrule height0.8pt}
}
}
}
$$
In this case, $|\delta|_{\mbox{\small val}}>1$.

\section{{ Genus Two Case }}

In this section we investigate genus two solutions of
the Toda equations using the hyperelliptic $\sigma$ functions
and $\psi$ functions.

The hyperelliptic $\sigma$ function was defined by Klein after
the prototype had been discovered by Weierstrass
\cite{Baker1, Klein, Weierstrass}.
Let us fix a hyperelliptic curve with genus two,
\begin{equation}
C_2\ :\ y^2= x^5 + \lambda_4 x^4 + \lambda_3 x^3 + \lambda_2 x^2
+ \lambda_1 x + \lambda_0,\label{eq:C2}
\end{equation}
where $\lambda_i$, $i=0,1,\cdots,4$ are complex numbers.
For a point in the symmetric product space of the curve $C_2$,
$((x_1,y_1),(x_2,y_2))\in \mbox{Sym}^2 C_2$,
its corresponding point $u\equiv(u_1,u_2)$ in the
Jacobi variety $J_2$ is given by
\begin{equation}
u_1:=\int^{(x_1,y_1)}_\infty \frac{d x}{y}
+\int^{(x_2,y_2)}_\infty \frac{d x}{y}, \quad
u_2:=\int^{(x_1,y_1)}_\infty \frac{x d x}{y}
+\int^{(x_2,y_2)}_\infty \frac{x d x}{y}.
\end{equation}
Here $\infty$ means the infinity point of the curve $C_2$.
At the point,
$\wp$ functions of genus two are defined as
\begin{equation}
\wp_{11}=\frac{f(x_1,x_2)-2 y_1y_2}{(x_1-x_2)^2},
\qquad \wp_{12} = x_1 x_2, \qquad \wp_{22} = x_1 +x_2,
\label{eq:wp2}
\end{equation}
where
$f(x,z):=\sum_{j=0}^{2}x^jz^j(\lambda_{2j+1}(x+z)+2\lambda_{2j})$.
It is known that there is a global function over ${\bf C}^2$
such that
\begin{equation}
\wp_{i j } =- \frac{\partial^2}{\partial u_i \partial u_j}
\log \sigma,
\end{equation}
which is the $\sigma$ function of genus two.

\subsection{{ Continuous Toda Equation}}

Though there were found solutions of the continuous Toda equation
in terms of the $\theta$ function related to a hyperelliptic
curve of genus $g$ in \cite{DateTanaka},
in this article we give another type of expression
of solutions in terms of $\sigma$ functions related to a
curve with genus two.

The additive formula of $\sigma$ function of genus two is
given by \cite{Baker2},
\begin{equation}
\frac{\sigma(v+u)\sigma(v-u)}{[\sigma(v)\sigma(u)]^2}
=-(\wp_{11}(u)-\wp_{11}(v) + \wp_{12}(u)\wp_{22}(v) -
\wp_{12}(v)\wp_{22}(u)).
\label{eq:add2}
\end{equation}
By letting
\begin{equation}
Q(u,v) := -(\wp_{11}(u)-\wp_{11}(v) + \wp_{12}(u)\wp_{22}(v) -
\wp_{12}(v)\wp_{22}(u)),
\end{equation}
we have
\begin{equation}
- \frac{\partial^2}{\partial u_i\partial u_j} \log (Q(u,v))
= \wp_{i j }(u+v) - 2\wp_{i j }(u) + \wp_{i j }(u-v).
\end{equation}
Let us fix
$u=n u_0+t$, $v=u_0$, constant numbers $b_{ij}:=\wp_{ij}(u_0)$
and
\begin{equation}
\Delta := \frac{\partial^2}{\partial t_1\partial t_1}
+b_{22}\frac{\partial^2}{\partial t_1\partial t_2}
+b_{12}\frac{\partial^2}{\partial t_2\partial t_2}.
\end{equation}
Then we have a relation,
\begin{equation}
Q((n+1) u_0+t,u_0)-2Q(n u_0+t, u_0)+Q((n-1) u_0+t,u_0)
=-\Delta \log Q(n u_0+t,u_0).
\end{equation}
By considering the relations (\ref{eq:wp2}) we let
$u_0$ correspond to a point
$((\bar x_1,\bar y_1),(\bar x_2,\bar y_2))\in \mbox{Sym}^2 C_2$
and then have
\begin{equation}
b_{22} = \bar x_1+\bar x_2,\quad
b_{12} = \bar x_1 \bar x_2.
\end{equation}
If the points are mutually conjugate or identical, {\it i.e.},
$\bar x_1\equiv\bar x_2$,
\begin{equation}
\Delta = ( \frac{\partial}{\partial t_1}
+\bar x_1\frac{\partial}{\partial t_2})^2.
\end{equation}
Hence for $t :=t_1 +t_2/\bar x_1$,
\begin{equation}
q_n := \log Q(n c+t,c),
\end{equation}
obeys the continuous Toda equation,
\begin{equation}
-\frac{d^2}{d t^2}q_n = \ee^{q_{n+1}}-2 \ee^{q_n} +\ee^{q_{n-1}}.
\end{equation}
As we showed in the genus one case,
this genus two solution also comes from the
addition formula (\ref{eq:add2}).

\subsection{{ Discrete Toda Equation}}

We give relations between the
discrete Toda equation and $\psi$ functions of genus two
as follows.
Generalizations of the $\psi$ function in (\ref{eq:psi1}) to
genus two curves are given by two different definitions;
one is defined over the Jacobi variety $J_2$
and another is defined over the curve $C_2$.
The former is studied by Kanayama \cite{kana}
and the latter is
investigated by Grant, Cantor, \^Onishi and this author
(see the references in \cite{matsu2}).
The definition by Kanayama is \cite{kana}
\begin{equation}
\psi_n(u)=
\frac{\sigma(nu)}{\sigma(u)^{n^2}} .\label{eq:psi2}
\end{equation}
Further he showed that $\psi_n$ obeys the same recursion relation
as (\ref{eq:rec_psi1}), viz
\begin{equation}
\psi_{n+m} \psi_{m-n}
=\left| \begin{array}{r r}
\psi_{m-1}\psi_n & \psi_{m}\psi_{n+1} \\
\psi_{m}\psi_{n-1} & \psi_{m+1}\psi_{n}
\end{array} \right| ,\label{eq:rec_psi2}
\end{equation}
basically using the the addition formula (\ref{eq:add2}).
Hence $\psi_k$ obeys a relation which has the same form as
(\ref{eq:rec4}). Kanayama gave the explicit forms of
$\psi_1$, $\psi_2$, $\psi_3$ and $\psi_4$ in terms of
$\wp$ functions (\ref{eq:wp2}) in \cite{kana}.
We can compute an explicit form of any $\psi_n$
as a rational function of
the affine coordinates $(x_1,y_2)$ and $(x_2,y_2)$
of the curves $\mbox{Sym}^2 C_2$ even though it is
too large to give its
explicit form here.

Due to its form, it is obvious that
(\ref{eq:rec_psi2}) is also related to the discrete Toda equation.
For mutually prime integers $p, q$ and an integer $n_0$,
we define quantities,
\begin{equation}
\phi_i^{\ j} := \psi_{n_0 + pi +q j},\label{eq:phi2}
\end{equation}
$\delta:=\psi_q/\psi_p$ and
$c(1-\delta^{2})=\psi_{p+q}\psi_{p-q}/\psi_p^2$. Then
(\ref{eq:rec_psi2}) becomes
\begin{equation}
\delta^{-2}\phi_i^{\ j+1}\phi_i^{\ j-1}
+c(1-\delta^{-2}) \phi_i^{\ j}\phi_i^{\ j}
-\phi_{i+1}^{\ \ j}\phi_{i-1}^{\ \ j}=0.
\label{eq:dtoda3}
\end{equation}
Hence we have a solution of
the discrete Toda equation (\ref{eq:dtodaV})
in \cite{hirota}
as shown in Section 2.2.

As a simple Abel variety has a division field
as its endomorphism in the category of the Abel variety
as it is known as Poincar\'e's complete reducibility theorem
\cite{lang}.
Hence, even though the Jacobi variety $J_2$ is two-dimensional,
an isometry $\varphi: J_2\to J_2$ is characterized by
an integer. The zeros of $\psi_n$ belonging to
${\bold Z}/n{\bold Z}$ determine the isometry.
Thus, as long as we deal with isometries of Jacobi variety,
an extension of the $\psi_n$ functions to
functions with double-index
must fail. It implies that
(\ref{eq:phi2}) is a natural in the sense of
a realization of the discrete equation
in category of the Abel variety.

\subsection{{Ultradiscrete Toda Equation}}

We consider the Jacobi variety $J_2$ as
a commutative ring
and its localization ring $R_{u_0}$ and/or
a field $K_{u_0}$ related to $R_{u_0}$.
Similar to the case of genus one, we
deal with a point of curve satisfying
\begin{equation}
c(1-\delta^2) =\frac{\psi_{p+q}(u)\psi_{p-q}(u)}
{\psi_{p}(u)^2}\in R_{u}^\times.
\end{equation}
By letting
\begin{equation}
f_i^{\ j}:=-\mbox{val}
\left(\frac{\phi_{i}^{\ \ j+1}\phi_{i}^{\ \ j-1}}
{\phi_{i}^{\ j}\phi_{i}^{\ j}} \right), \quad
d :=- \mbox{val}(\delta^2),
\end{equation}
and being supposed that for all of $i$ and $j$,
the valuations of the additions of $f$ s expressed by
the minimal functions like the equality case
in the second relation of (\ref{eq:val2}),
we find a solution of
the ultradiscrete Toda equation in \cite{takahashi},
\begin{equation}
f_{i}^{\ j+1} - 2f_{i}^{\ j}+f_{i}^{\ j-1}
= \mbox{max}(0, f_{i+1}^{\ j}+d)-2
\mbox{max}(0, f_{i}^{\ j}+d)
+\mbox{max}(0,f_{i-1}^{\ j}+d ).
\end{equation}

Even though the case of genus one gives us
trivial solutions,
genus two case is expected to provide
us nontrivial solutions
because it has larger degree of freedom than the elliptic
curve case.

\vskip 1.0 cm
\section{{ Discussion }}
\vskip 0.5 cm

In this article we have considered the relations between
the Toda equations in the
continuous, discrete and ultradiscrete levels
and $\sigma$ functions of genera one and two.
We showed that these solutions, in principle, come from
the addition formulae of the algebraic functions
over algebraic curves of genera one and two.

As we started from the curves, all of the solutions are
expressed by points at curves (\ref{eq:ecurve}) and
(\ref{eq:C2}).
As we gave some explicit solutions related to elliptic curves,
we can basically find
explicit forms of the other solutions
in terms of points of the curves even of genus two,
though they might be slightly complicated.
As a next step, we should give more explicit computations
of the $\psi$ functions
on the genus two case by finding a nicer strategy to handle
the huge polynomials.
However, it is remarkable that in our solutions,
there do not appear excess parameters
except the coefficients $\lambda_i$, $i=0,1,\cdots,4$
in (\ref{eq:ecurve}) and
(\ref{eq:C2}). In other words we have no ambiguity for
the parameters even in genus two case and
it means that we are free from the so-called
Schottky problem. This contrasts with
the solutions in terms of the $\theta$ functions
over the Jacobi variety \cite{DateTanaka, tokihiro}.
Of course as it might be difficult to deal with
the hyperelliptic integrals, thus we should
select the methods according to the circumstances.

Further it is interesting that the ultradiscrete equations
can be defined on the Jacobi varieties associated with
nondegenerate algebraic curves over a field with
character zero using the concept of discrete valuation.
(In \cite{matsu3} we show that the ultradiscrete equations can
be defined over fields with nonvanishing character.)
It means that, if we find a recursion relation over an algebraic
variety, we might have its ultradiscrete version by
taking its discrete valuation.

Finally we comment on the higher genus case. Unfortunately
since the addition formula is not simple \cite{Baker1, BEL},
we could not deal with
$\sigma$ functions associated with a curve with a higher genus
as mentioned above.
We hope that we can obtain
such solutions in future. We note that
the paper \cite{BEL} may have
some effects on the study.

\vskip 1.0 cm
{\Large \bf{Acknowledgement }}
\vskip 0.5 cm

I am grateful to Prof. Tokihiro for sending me
his interesting paper with Kimijima \cite{tokihiro}.
I thank to Prof. \^Onishi for directing me to the theory
of Poincar\'e in \cite{lang}.

\vskip 1.0 cm
\appendix
{\Large \bf{Appendix }}
\vskip 0.5 cm
\setcounter{section}{1}

Let us deal with $y^2 = x^3 + 1/4$, $y^2 = x^3 -x$ and
$y^2 = x^2(x + 1/4)$ and show explicit function
forms of their $\psi$ functions.

\subsection{{ $y^2 = x^3 + 1/4$}}

\begin{eqnarray}
\psi_1 &=& 1,
\\
\psi_2 &=& - 2y,
\\
\psi_3 &=& 3 x (1 + x) (1 - x + x^2) = 3 x (1+x^3),
\\
\psi_4 &=&\psi_2(-1 + 10 x^3 + 2 x^6),
\\
\psi_5 &=&-1 - 25 x^3 - 15 x^6 + 95 x^9 + 5 x^{12},
\\
\psi_6 &=& \psi_2 \psi_3 (-2 + x^3)
(1 - 3 x + 3 x^2 + x^3) \nonumber\\
&&
\times (1 + 3 x + 6 x^2 + 11 x^3 + 12 x^4 - 3 x^5 + x^6),
\\
\psi_7 &=& (1 - x^3 + 7 x^6) \nonumber\\
&&
\times (1 - 48 x^3 - 741 x^6 - 1924 x^9 - 363 x^{12} + 141 x^{15} +
x^{18}),
\\
\psi_8 &=&\psi_4
(-1 - 104 x^3 - 952 x^6 - 4124 x^9 - 3430 x^{12}
\nonumber \\
&&-
1544 x^{15}- 7336 x^{18}+ 616 x^{21}+ 2 x^{24}),\\
\psi_9 &=& 3 \psi_3(1 - 3 x^2 + x^3)
(1 + 3 x^2 + 2 x^3 + 9 x^4 + 3 x^5 + x^6) \nonumber \\
&&
\times(1 + 9 x^2 + 3 x^3 + 18 x^5 - 24 x^6 + 9 x^8 + x^9) \nonumber \\
&&\times
(1 - 9 x^2 + 6 x^3 + 81 x^4 - 45 x^5 - 39 x^6 + 324 x^7 \nonumber \\
&&\quad
+ 153 x^8 - 142 x^9 + 486 x^{10}+ 396 x^{11}+ 582 x^{12}\nonumber \\
&&\quad+
324 x^{13}+ 198 x^{14}- 48 x^{15}+ 81 x^{16}- 9 x^{17}+ x^{18}),\\
\psi_{10} &=& \frac{1}{2} \psi_2 \psi_5
(1 - 177 x^3 - 474 x^6 - 7070 x^9 - 104805 x^{12} \nonumber \\
&&\quad- 542232 x^{15}- 862941 x^{18}- 1404072 x^{21}- 368055 x^{24}
\nonumber \\
&&\quad+
29380 x^{27}- 55284 x^{30}+ 1173 x^{33} + x^{36}),\\
\psi_{11} &=&
-1 - 242 x^3 + 605 x^6 + 102729 x^9 + 2270301 x^{12} \nonumber \\
&&\quad+
17393277 x^{15}+ 59389374 x^{18}+ 189881835 x^{21}\nonumber \\
&&\quad
+1106263389 x^{24}+ 4869514969 x^{27}+ 10595519759 x^{30} \nonumber \\
&&\quad+
8054721004 x^{33} - 22319781 x^{36}- 4760052033 x^{39} \nonumber \\
&&\quad-
8579472693 x^{42 }- 1596123771 x^{45} + 66133914 x^{48} \nonumber \\
&&\quad-
62045313 x^{51} - 1153603 x^{54} + 23221 x^{57} + 11 x^{60},
\\
\psi_{12}
&=&\psi_3\psi_4 (-2 + x^3)
(1 - 3 x + 3 x^2 + x^3) \nonumber \\
&&\times
(1 + 3 x + 6 x^2 + 11 x^3 + 12 x^4 - 3 x^5 + x^6) \nonumber \\
&&\times
(-2 - 32 x^3 - 84 x^6 - 134 x^9 + x^{12})
\nonumber
\\
&&\times
(1 + 6 x + 12 x^2 + 4 x^3 + 45 x^4 + 36 x^5 + 60 x^6
\nonumber \\
&&\quad+
72 x^7 - 45 x^8 + 58 x^9 - 48 x^{10}+ 12 x^{11}+ x^{12})
\nonumber
\end{eqnarray}
\begin{eqnarray}
&&\times
(1 - 6 x + 24 x^2 - 64 x^3 + 75 x^4 + 456 x^5 - 620 x^6
\nonumber \\
&& \quad+
252 x^7 + 2070 x^8 - 1618 x^9 - 3072 x^{10}+ 3216 x^{11}
\nonumber
\\
&&\quad+
4003 x^{12}- 9696 x^{13}+ 1416 x^{14}+ 11396 x^{15}
\nonumber \\
&&\quad+
1548 x^{16}- 5058 x^{17}+ 460 x^{18}+ 1632 x^{19}
\nonumber \\
&&\quad+ 1653 x^{20}+
692 x^{21}+ 192 x^{22}- 12 x^{23}+ x^{24}),
\\
\psi_{13}&=&
(1 + 16 x^3 + 96 x^6 + 13 x^9 + 13 x^{12})
\nonumber \\
&&\times
(1 - 354 x^3 - 17247 x^6 + 92420 x^9 - 6264417 x^{12}
\nonumber \\
&&\quad
-91630974 x^{15}- 414038735 x^{18}- 631690011 x^{21}
\nonumber \\
&&\quad+
3596512338 x^{24}+ 43118516972 x^{27}+ 215967505719 x^{30}
\nonumber \\
&&\quad+
533661527514 x^{33} + 582732421153 x^{36}+ 284118813696 x^{39 }
\nonumber
\\
&&\quad+
450924775284 x^{42} + 1313707269872 x^{45} + 1846766455056 x^{48}
\nonumber \\
&& \quad+
403474854555 x^{51} - 263110973327 x^{54} - 22534762701 x^{57}
\nonumber
\\
&&\quad+
685417938 x^{60} - 111537892 x^{63} - 798438 x^{66}
\nonumber \\
&&\quad
+ 5748 x^{69} +
x^{72}),\\
\psi_{14} &=&\psi_2 \psi_7
(1 - 48 x^3 - 741 x^6 - 1924 x^9 - 363 x^{12}+
141 x^{15}+ x^{18}) \nonumber \\
&&\times
(1 + 504 x^3 + 2421 x^6 + 5676 x^9 + 166356 x^{12}
\nonumber \\
&&\quad+
3098475 x^{15}+ 22597638 x^{18}+ 56826270 x^{21}
\nonumber \\
&&\quad-
73281168 x^{24}- 582904249 x^{27}- 862862121 x^{30}\nonumber \\
&&\quad+
133470252 x^{33} + 317907519 x^{36}- 632536713 x^{39}
\nonumber \\
&& \quad-
77646699 x^{42} - 41502855 x^{45 }- 2997252 x^{48}
\nonumber \\
&& \quad + 8847 x^{51} +
x^{54}),
\\
\psi_{15}
&=&\psi_3\psi_5
(-5 + 65 x^3 + 685 x^6 + 3410 x^9 + 11425 x^{12}
\nonumber \\
&&\quad+
5735 x^{15}+ 3145 x^{18}- 520 x^{21}+ x^{24})\nonumber \\
&&\times
(1 - 6 x + 6 x^2 + 44 x^3 + 21 x^4 - 21 x^5 + 676 x^6
\nonumber \\
&&\quad+
9 x^7 - 9 x^8 + 569 x^9 + 2841 x^{10}- 2841 x^{11}\nonumber \\
&&\quad-
1694 x^{12}+ 13119 x^{13}- 13119 x^{14}+ 10019 x^{15}
\nonumber \\
&&\quad-
4284 x^{16}+ 4284 x^{17}+ 4591 x^{18}- 1446 x^{19}+ 1446 x^{20}
\nonumber \\
&&\quad-
496 x^{21}- 24 x^{22}+ 24 x^{23}+ x^{24})
\nonumber \\
&&\times
(1 + 6 x + 30 x^2 + 124 x^3 + 279 x^4 - 495 x^5 + 3036 x^6
\nonumber \\
&&\quad +
2871 x^7 - 2790 x^8 + 60959 x^9 - 13686 x^{10}- 19695 x^{11}
\nonumber \\
&&\quad+
469946 x^{12}- 200034 x^{13}+ 128295 x^{14}+ 602229 x^{15}
\nonumber \\
&&\quad-
2440926 x^{16}+ 3056445 x^{17}- 422129 x^{18}- 9809094 x^{19}
\nonumber \\
&&\quad+
18607485 x^{20}+ 20165779 x^{21}+ 8262864 x^{22}
\nonumber \\
&&\quad+
74286585 x^{23}+ 94839246 x^{24}+ 71549460 x^{25}
\nonumber \\
&&\quad+
150594579 x^{26} + 118349119 x^{27}- 3156510 x^{28}
\nonumber \\
&&\quad+
30275751 x^{29} - 36357239 x^{30}- 138954870 x^{31}
\nonumber \\
&&\quad -
1389186 x^{32} + 73952184 x^{33} + 20894985 x^{34}
\nonumber \\
&&\quad-
28859229 x^{35} + 22894661 x^{36}+ 16500675 x^{37}
\nonumber \\
&&\quad -
2444511 x^{38} - 2237686 x^{39} + 1693800 x^{40}
\nonumber \\
&&\quad + 672156 x^{41} +
324606 x^{42} + 58950 x^{43} + 11034 x^{44}
\nonumber \\
&&\quad- 416 x^{45} +
600 x^{46} - 24 x^{47} + x^{48}).
\end{eqnarray}

\subsection{{ $y^2 = x(x^2 - 1)$}}

\begin{eqnarray}
\psi_1 &=& 1,
\\
\psi_2 &=& -2 y,
\\
\psi_3 &=&3 (-2 + x) x^2 (2 + x),
\\
\psi_4 &=& -4y\ x^2 (-6 - 15 x^2 + x^4),
\\
\psi_5 &=& x^4 (-192 + 1632 x^2 - 496 x^4 - 220 x^6 + 5 x^8),
\end{eqnarray}
\begin{eqnarray}
\psi_6 &=& - 6y \ (-2 + x) x^6 (2 + x)
\nonumber \\ &&
\times(-336 + 912 x^2 - 1348 x^4 - 100 x^6 + x^8),
\\
\psi_7 &=& x^8 (
27648 + 483840 x^2 - 2951424 x^4 + 2595456 x^6 - 1101888 x^8
\nonumber \\ && \quad +
447840 x^{10} - 31376 x^{12} - 1544 x^{14} + 7 x^{16}),
\\
\psi_8 &=& -8y\ x^{10} (-6 - 15 x^2 + x^4)
\nonumber \\ &&\times
(-18432 + 603648 x^2 - 2432640 x^4 + 2577312 x^6 -
\nonumber \\ && \quad
702392 x^8 + 47744 x^{10} - 23070 x^{12} - 412 x^{14} + x^{16}),
\\
\psi_9 &=&
-3(-2 + x) x^{14}(2 + x)
\nonumber \\ &&\times
(16367616 - 154607616 x^2 + 1527054336 x^4 - 5301780480 x^6
\nonumber \\ &&
+ 4162000896 x^8 + 567207936 x^{10} - 1938695936 x^{12}
+ 731321472 x^{14}
\nonumber \\ &&
- 1489472 x^{16} + 5367072 x^{18}
- 164000 x^{20}
- 2316 x^{22} + 3 x^{24}) .
\end{eqnarray}

\subsection{{ $y^2 = x^2(x + 1/4)$}}

\begin{eqnarray}
\psi_1 &=& 1,
\\
\psi_2 &=& -2 y,
\\
\psi_3 &=& x^3 (1 + 3 x),
\\
\psi_4 &=& -2y \ x^5 (1 + 2 x),
\\
\psi_5 &=& x^{10} (1 + 5 x + 5 x^2),
\\
\psi_6 &=& -2y \ x^{14} (1 + x) (1 + 3 x),
\\
\psi_7 &=& x^{21} (1 + 7 x + 14 x^2 + 7 x^3),
\\
\psi_8 &=& -2 y\ x^{27} (1 + 2 x) (1 + 4 x + 2 x^2),
\\
\psi_9 &=& x^{36} (1 + 3 x) (1 + 6 x + 9 x^2 + 3 x^3),
\\
\psi_{10} &=& -2y\ x^{44} (1 + 3 x + x^2) (1 + 5 x + 5 x^2),
\\
\psi_{11} &=& x^{55} (1 + 11 x + 44 x^2 + 77 x^3 + 55 x^4 + 11 x^5),
\\
\psi_{12} &=& -2y \ x^{65} (1 + x) (1 + 2 x) (1 + 3 x)
(1 + 4 x + x^2),
\\
\psi_{13} &=& x^{78}
(1 + 13 x + 65 x^2 + 156 x^3 + 182 x^4 + 91 x^5 + 13 x^6),
\\
\psi_{14} &=& -2y\ x^{90} (1 + 5 x + 6 x^2 + x^3)
(1 + 7 x + 14 x^2 + 7 x^3),
\\
\psi_{15} &=&
x^{105}(1 + 3 x)(1 + 5 x + 5 x^2)
(1 + 7 x + 14 x^2 + 8 x^3 + x^4)\\
\psi_{16} &=&- 2 y\ x^{119} (1 + 2 x) (1 + 4 x + 2 x^2)
(1 + 8 x + 20 x^2 + 16 x^3 + 2 x^4).
\end{eqnarray}

\end{document}